\documentclass[twocolumn,showpacs,preprintnumbers,amsmath,amssymb]{revtex4}

\usepackage{graphicx}
\usepackage{dcolumn}
\usepackage{bm}

\def\btbl{\begin{tabular}} \def\etbl{\end{tabular}}
\def\bcc{\begin{center}} \def\ecc{\end{center}}
\def\beq{\begin{equation}} \def\eeq{\end{equation}}
\def\btbl{\begin{tabular}} \def\etbl{\end{tabular}}

\def\E941{{\footnotesize E941}} 
\def\NA49{{\footnotesize NA49}} \def\NA35{{\footnotesize NA35}}

\begin{document}
\title{Strange quark suppression and strange hadron production\\
       in pp collisions at RHIC and LHC}
\author{Hai-Yan Long}
\affiliation{Dept. of Physics, College of Science, China Three Gorges Univ.,
Yichang 443002, China}
\author{Sheng-Qin Feng}
\email{fengsq@ctgu.edu.cn}
\affiliation{Dept. of Physics, College of Science, China Three Gorges Univ.,
Yichang 443002, China}
\author{Dai-Mei Zhou}
\affiliation{Institute of Particle Physics, Huazhong Normal University, Wuhan
 430082, China}
\author{Yu-Liang Yan}
\affiliation{China Institute of Atomic Energy, P. O. Box 275 (18), Beijing
102413, China}
\author{Hai-Liang Ma}
\affiliation{China Institute of Atomic Energy, P. O. Box 275 (18), Beijing
102413, China}
\author{Ben-Hao Sa}
\email{sabh@ciae.ac.cn}
\affiliation{China Institute of Atomic Energy, P. O. Box 275 (18), Beijing
102413, China}
\affiliation{Institute of Particle Physics, Huazhong Normal University, Wuhan
430082, China}
\affiliation{~CCAST (World Laboratory), P. O. Box 8730, Beijing 100080, China}
\begin{abstract}
The parton and hadron cascade model PACIAE based on PYTHIA is utilized
to systematically investigate strange particle production in pp collisions 
at the RHIC and LHC energies. Globally speaking, the PACIAE results
of the strange particle rapidity density at mid-rapidity and the transverse
momentum distribution are better than PYTHIA (default) in comparing with
STAR and ALICE experimental data. This may represent the importance of the
parton and hadron rescatterings, as well as the reduction mechanism of 
strange quark suppression, added in the PACIAE model. The $K/\pi$ ratios as 
a function of reaction energy in pp collisions from SPS to LHC energies are 
also analyzed in this paper. \\

\vskip0.05cm \noindent Keywords: strange particle, rapidity distribution,
transverse momentum distribution, PYTHIA model, PACIAE model.
\end{abstract}

\pacs{25.75.Dw, 24.10.Lx}

\maketitle

\section{Introduction}
\label{intro}
Strange particle production is a powerful probe into the hadronic interaction
and the hadronization process in pp and heavy ion collisions at relativistic
energies. The strangeness enhancement in relativistic nucleus-nucleus
collisions relative to pp collisions at the same energy has been proposed
as a signature of the Quark-Gluon Plasma (QGP) formation in the relativistic
heavy ion collisions \cite{rafe}. This is based on the principle that the
threshold energy of strange particle production in Quark-Gluon Matter (QGM)
is higher than that in hadronic matter. Unfortunately, the QGM has not yet
been confirmed to be unique explanation of strangeness enhancement.

The Large Hadron Collider (LHC) at CERN opens a new era in the
investigation of nucleus-nucleus collisions at relativistic energy. It has
established a new platform to study the properties of QCD matter (QGM)
\cite{pop,arme}. Strangeness production is one of the most important research 
topics in the relativistic nuclear collisions at the LHC. Recently, the
ALICE collaboration reported their data of strange particle production in pp
collisions at $\sqrt{s}= 0.9$ TeV~\cite{alice1,alice2}. This introduces a new
energy regime in the research of strangeness production and the possibility 
to compare with the previous measurements in pp collisions at $\sqrt{s}$=200
GeV \cite{star1}.

In the LUND string fragmentation scheme \cite{ander},  the suppression of $s$
quark pair production compared with $u$ ($d$) pair production (the parameter
$parj(2)$ in PYTHIA or $\lambda$ denoted later) was assumed to be fixed.
However, later experiments \cite{ua1} have shown that this suppression
decreases with increasing reaction energy. In Ref.~\cite{sa4}, the reduction
mechanism of the strange quark suppression has been introduced in the LUCIAE
model. By this mechanism, they described the strangeness enhancement in pp,
p+A, and A+A collisions at SPS energies successfully \cite{sa4,sa5}. LUCIAE
is a hadron cascade model based on FRITIOF \cite{pi} with the firecracker
model and hadronic rescattering added \cite{sa1}.

In this paper, the reduction mechanism of the strange quark suppression was 
introduced in the parton and hadron cascade model PACIAE \cite{sa2} and  
denoted as the modified PACIAE model to distinguish it from the default one. 
The modified PACIAE model was then used to systematically investigate the 
strange particle production in pp collisions at the RHIC and LHC energies.  

This paper is organized as follows. In Sec. II, we give a brief review the 
PACIAE model \cite{sa2} and the reduction mechanism of strangeness quark 
suppression \cite{sa4}. In Sec. III, we use the modified parton and hadron 
cascade model PACIAE based on PYTHIA \cite{sjo} to analyze systematically
the strangeness production in pp collisions at RHIC and LHC energies 
\cite{star1,alice1,alice2}. The $K/\pi$ ratio as a function of reaction 
energy in pp collisions at relativistic energies from SPS to LHC was also 
investigated in Sec.III. Section IV gives a summary and conclusion.

\section{Models}
PACIAE \cite{sa2} is a parton and hadron cascade model based on PYTHIA
\cite{sjo} which is a model for high energy hadron-hadron (hh) collisions
at the hadronic level. In the PYTHIA model, a pp collision is decomposed into
parton-parton collisions. The hard parton-parton interaction is described by
the lowest leading order perturbative QCD (LO-pQCD). The soft parton-parton
collision, a non-perturbative phenomenon, is considered empirically. The
initial- and final-state QCD radiations, as well as the multiparton
interactions, are considered. Therefore the consequence of a pp collision is
a parton multijet configuration composed of quarks (anti-quarks), di-quarks
(anti-diquarks) and gluons, along with a few hadronic remnants. After that, 
the string construction and fragmentation (hadronization) are preformed to 
obtain a hadronic final state for a pp collision.

For the pp collision, the PACIAE model is different from PYTHIA in the 
addition of the parton initiation stage, the parton rescattering before
hadronization, and the hadron rescattering after hadronization. Thus, the
PACIAE model consists of the parton initiation, parton evolution
(rescattering), hadronization, and hadron evolution (rescattering) four
stages.

In order to create the parton initiation stage for a pp collision, the string
fragmentation is switched off temporarily in PACIAE. The di-quarks
(anti-diquarks) are broken up. Then the partonic initial state is obtained 
in the PACIAE model, instead of the hadronic final state in PYTHIA. This is
just QGM formed in the parton initiation stage of a pp collision.

The rescattering among partons in QGM is considered by the $2\rightarrow2$ LO
pQCD parton-parton cross sections \cite{plb70}. By integrating this 
differential cross section properly, the total cross section is obtained. 
With the differential and total cross sections, parton rescattering is 
performed by the Monte Carlo method.

In the hadronization stage, the QGM formed after parton rescattering is
hadronized by the LUND string fragmentation regime \cite{ander,sjo} or the
Monte Carlo coalescence model proposed in \cite{sa2}. The LUND string
fragmentation is used in this paper.

The hadronic rescattering is modeled with the usual two body elastic and/or 
inelastic collisions \cite{sa1}, until the hh collision pairs are exhausted 
(hadronic freeze-out). The rescatterings among $\pi,~K,~p,~n,~\rho(\omega),
~\Delta,~\Lambda,~\Sigma,~\Xi,~J/\Psi$ and their antiparticles are considered 
for the moment.

In the PYTHIA model \cite{sjo}, it is assumed that the $q\bar{q}$
pair with quark mass $m$ and transverse momentum $p_T$ is 
first created quantum mechanically at one point and then tunnel
out to the classically allowed region. This tunnelling probability
is calculated 
\begin{equation}
\exp({-\frac{{\pi}m^2}{\kappa}})\exp({-\frac{{\pi}p^2_T}{\kappa}}),
\label{tunn}
\end{equation}
where the string tension is assumed to be a constant of $\kappa\approx$1
GeV/fm$\approx 0.2$ GeV$^2$ \cite{ander,sjo}. This probability implies a
suppression of strange (heavy) quark production: $u:d:s:c\approx
1:1:0.3:10^{-11}$. Therefore the charm and heavier quarks are not expected
to be produced in the soft string fragmentation process but only in the
hard process or as a part of the initial- and final-state QCD radiations.

A reduction mechanism of strange quark suppression was introduced
in \cite{sa4} by assuming that the effective string tension increased
with increasing reaction energy. Hence the strange (heavy) quark production 
increased with increasing reaction energy. It was further assumed in
\cite{sa4} that the effective string tension variation with reaction
energy could be considered by the effective string tension as a function of 
the number and the hardening of gluons in a single string as follows:
\begin{equation}
\kappa^{eff}=\kappa_0(1-\xi)^{-\alpha},
\end{equation}
\begin{equation}
\xi=\frac{\ln(\frac{k^2_{Tmax}}{s_0})}{\ln(\frac{s}{s_0})+\sum_{j=2}
^{n-1}\ln(\frac{k^2_{Tj}}{s_0})},
\label{xi}
\end{equation}
where $\kappa_0$ is the string tension of a pure $q\bar q$ string,  
assumed to be $\sim1$ GeV/fm. Here it should be mentioned
that the above Eq.~(\ref{xi}) represents the deviation scale of the
multigluon string from the pure string. The gluons in a
multigluon string are ordered from $2$ to $n-1$, because of the
quark and antiquark on both ends of string with index $1$ and $n$,
respectively. $k_{Tj}^{2}$ ($k_{Tmax}$) is the transverse momentum
of gluon $j$ with $k_{Tj}>s_0$ (gluon largest transverse
momentum). The parameters $\alpha$=3.5 GeV and $\sqrt{s_0}$=0.8
GeV were determined by fitting the hh collision data \cite{sa4}.

The strange quark suppression factor $\lambda$ (i.e. $parj(2)$ in PYTHIA) 
and the width $\sigma$ ($parj(21)$ in PYTHIA) of the Gaussian $p_x$ and $p_y$
transverse momentum distributions of the primary hadrons, in a string with
effective string tension $\kappa^{eff}_1$, are denoted by $\lambda_1$ and
$\sigma_1$, respectively. These two quantities in a string with effective
string tension $\kappa^{eff}_2$, $\lambda_2$ and $\sigma_2$, can be 
calculated by Eq.~\eqref{tunn}
\begin{equation}
 \lambda_2=\lambda_1^{\frac{\kappa^{eff}_1}{\kappa^{eff}_2}},
\label{lamd}
\end{equation}
\begin{equation}
\sigma_2=\sigma_1(\frac{\kappa^{eff}_2}{\kappa^{eff}_1})^{1/2}.
\end{equation}

In the PYTHIA model there are parameters parj(1) and parj(3) related to 
strangeness production, besides parj(2) and parj(21). parj(1) is 
the suppression of diquark-antidiquark pair production relative to the 
quark-antiquark pair production. parj(3) refers to the extra suppression 
of strange diquark production relative to the normal suppression of strange
quark production. For double strange particle (strange baryon) production, 
parj(1) and parj(3) have to be considered as well. It is not hard to prove 
that the Eq.~\eqref{lamd} is also valid for $parj(1)$ and $parj(3)$.

\section{Calculations and results}
The reduction mechanism of the strange quark suppression has been included in
the PACIAE model. One might first tune the parameters $parj(1)$, $parj(2)$,
and $parj(3)$ to fit the strangeness production data in a given nuclear
collision system at a given energy. The resulting $parj(1)$, $parj(2)$, and 
$parj(3)$ and the effective string tension can be used to predict the
strangeness production in the same reaction system at different energies,
even in different reaction systems.

As we aimed at the physics behind the strange particle production and not at
the reproduction of experimental data, we kept all model parameters at their
default values. However, the higher order and non-perturbative correction 
factor in parton-parton interaction, $K$, was assumed to be 3.

We then first globally tuned the parameters $parj(1)$, $parj(2)$, and  
$parj(3)$ in default PACIAE simulations to fit the strangeness production 
data from NSD (non single diffractive) pp collisions at $\sqrt s$=200 GeV 
\cite{star1}. The results are shown in Tab.~\ref{200} along with the 
results from default PYTHIA simulations. From this table, we find that 
the STAR data are better described by default PACIAE model rather than by the 
default PYTHIA model.

\begin{figure}[h!]
\centering \resizebox{0.43\textwidth}{!}{
\includegraphics{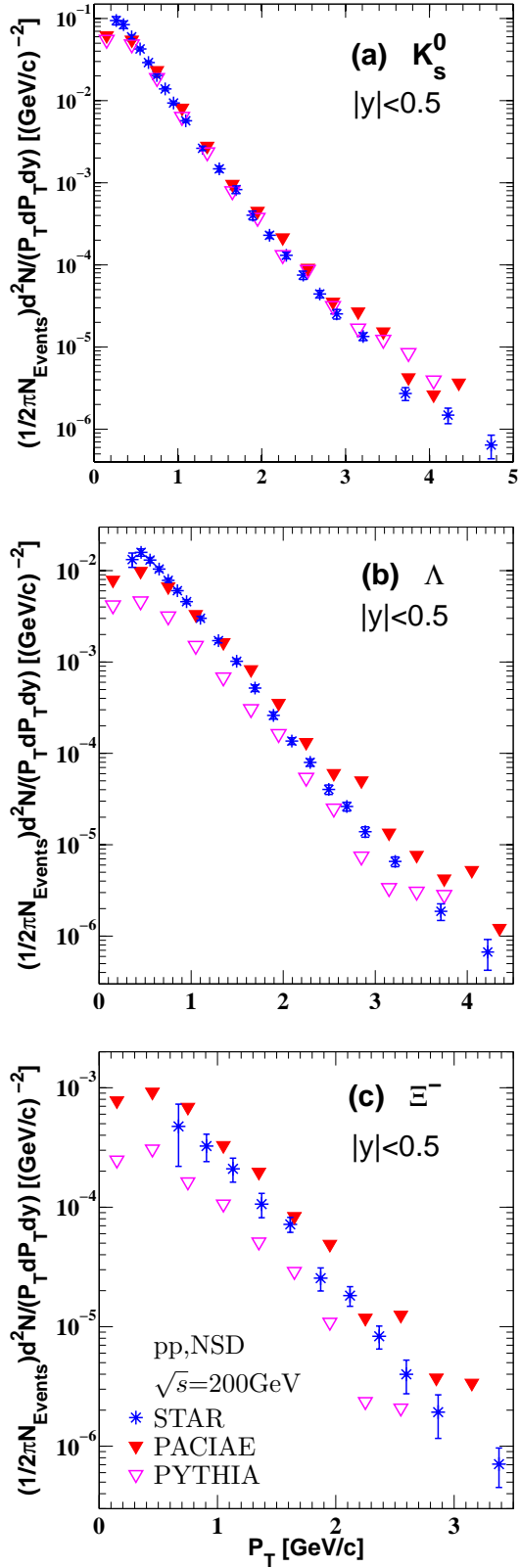}}
\caption{(Color online)~Transverse momentum distribution of
strange particles in NSD pp collisions at $\sqrt s$=200 GeV.
Panels (a), (b) and (c) are for $K_s^0$ , $\Lambda$, and  $\Xi^-$,
respectively. The STAR data were taken from \cite{star1}.}
\label{star}
\end{figure}

The strange particle transverse momentum spectra in the NSD pp collisions
at $\sqrt s$=200 GeV calculated by the default PACIAE and default PYTHIA 
are shown in Fig.~\ref{star}. The corresponding STAR data \cite{star1} are 
also given. The panels (a), (b), and (c) are for $K_s^0$, 
$\Lambda$, and $\Xi^-$, respectively. These results indicate that, globally 
speaking, the strange particle $p_T$ spectrum is better described by the 
default PACIAE rather than the default PYTHIA model. Here, the role of parton 
and hadron rescatterings is also not negligible.

The fitted parameters of $parj(1)$=0.18, $parj(2)$=0.4, and
$parj(3)$=0.5 as well as the calculated $\kappa^{eff}$=1.387 were then
used to calculate the strange particle production in INEL (inelastic) pp 
collisions at $\sqrt s$=900 GeV by the modified PACIAE model. The
resulting strange particle rapidity densities at mid-rapidity were
compared with the ALICE data \cite{alice1} as well as the results of default 
PYTHIA in Tab.~\ref{900}. One sees in this table that the ALICE data are 
better reproduced by modified PACIAE, rather than the default PYTHIA model. 
The effects of the reduction mechanism of strange quark suppression and the 
parton and hadron rescatterings may be all important which has to be 
separately studied.

\begin{figure*}[h!]
\centering \resizebox{0.85\textwidth}{!}{
\includegraphics{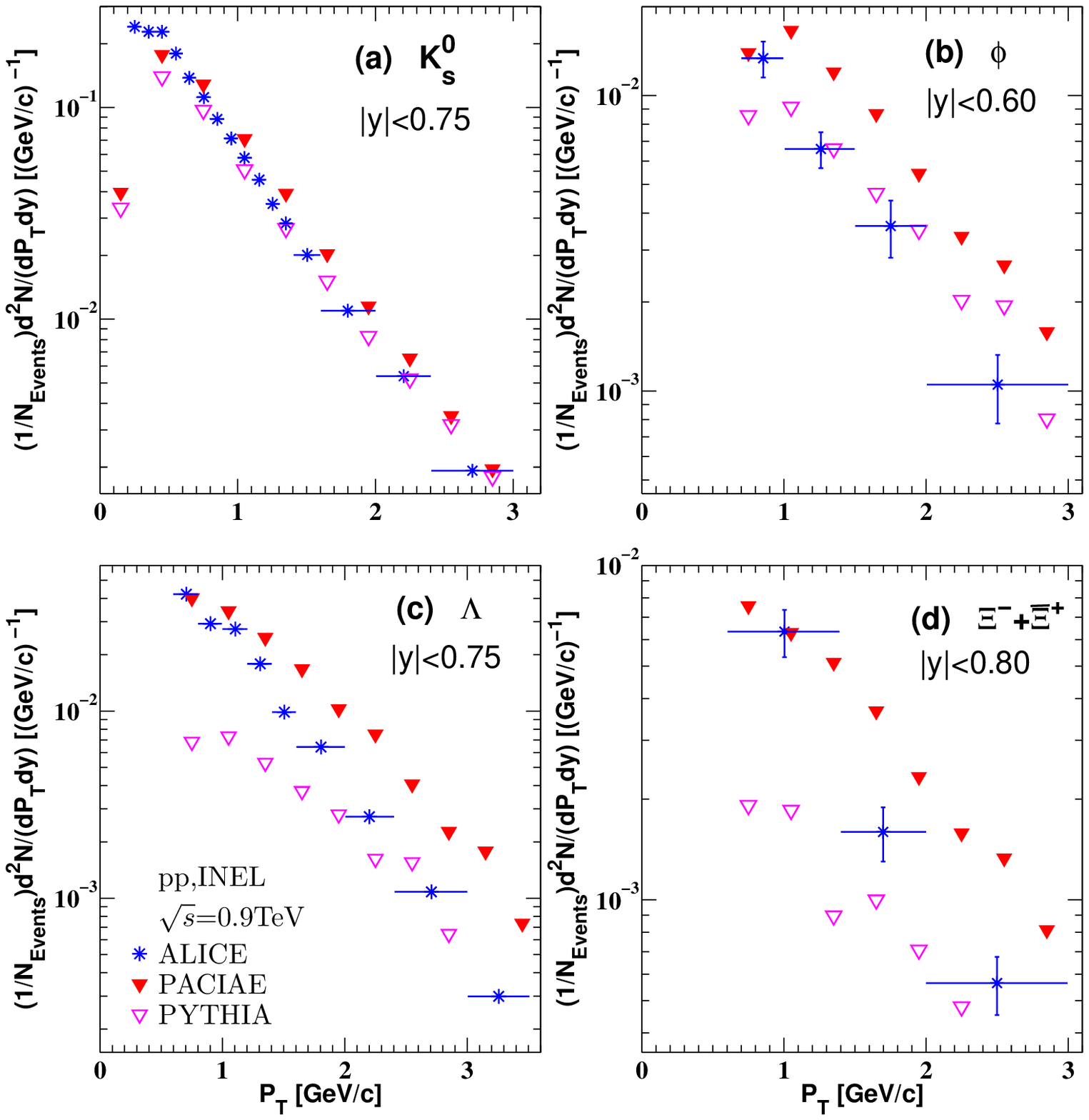}}
\caption{(Color online)~Transverse momentum distributions of
strange particles in INEL pp collisions at $\sqrt s$=900 GeV.
Panels (a), (b), (c), and (d) are for $K_{s}^0$, $\phi$,
$\Lambda$, and $\Xi^-$+$\overline\Xi^+$, respectively. The ALICE
data were taken from \cite{alice1}} \label{alice}
\end{figure*}

We present the strange particle transverse momentum distributions in INEL pp
collisions at $\sqrt s$=900 GeV calculated by the modified PACIAE and default 
PYTHIA models in Fig.~\ref{alice}. The corresponding ALICE data \cite{alice1} 
are also given. The panels (a), (b), (c), and (d) are for $K_{s}^0$, $\phi$, 
$\Lambda$, and $\Xi^-$+$\overline\Xi^+$, respectively. One sees in 
Fig.~\ref{alice} that the modified PACIAE results fit well with the ALICE 
data for the light strange particle $K_{s}^0$. However, for the $\phi$ meson 
and the heavy strange baryons ($\Lambda$ and $\Xi^-$+$\overline\Xi^+$) the 
modified PACIAE model overestimates the data in the $p_T>$1 GeV/c region. 
This may be because we only used the default PACIAE to tune the parameters 
to the data of strange particle rapidity densities at mid-rapidity in $\sqrt 
s$=200 GeV pp collisions \cite{star1}. In this sense, the transverse momentum 
distribution is calculated parameter free. Therefore, even if the calculated 
strange particle rapidity densities at mid-rapidity are close to the data 
(cf. Tab.~\ref{200} and Tab.~\ref{900}), the calculated transverse momentum 
distribution may not be closely similar to the data. Thus, the calculated 
transverse momentum distribution is softer than the data in the $p_T<$1 GeV/c 
region, but harder in the $p_T>$1 GeV/c region. It might also be possible 
that the reduction mechanism of strange quark suppression is not so suitable 
for the heavy strange baryons. Therefore, the role played by the reduction 
mechanism of strange quark suppression and the parton and hadron 
rescatterings are worthy to be separately studied in the next investigations.

\begin{figure}[h!]
\centering \resizebox{0.43\textwidth}{!}{
\includegraphics{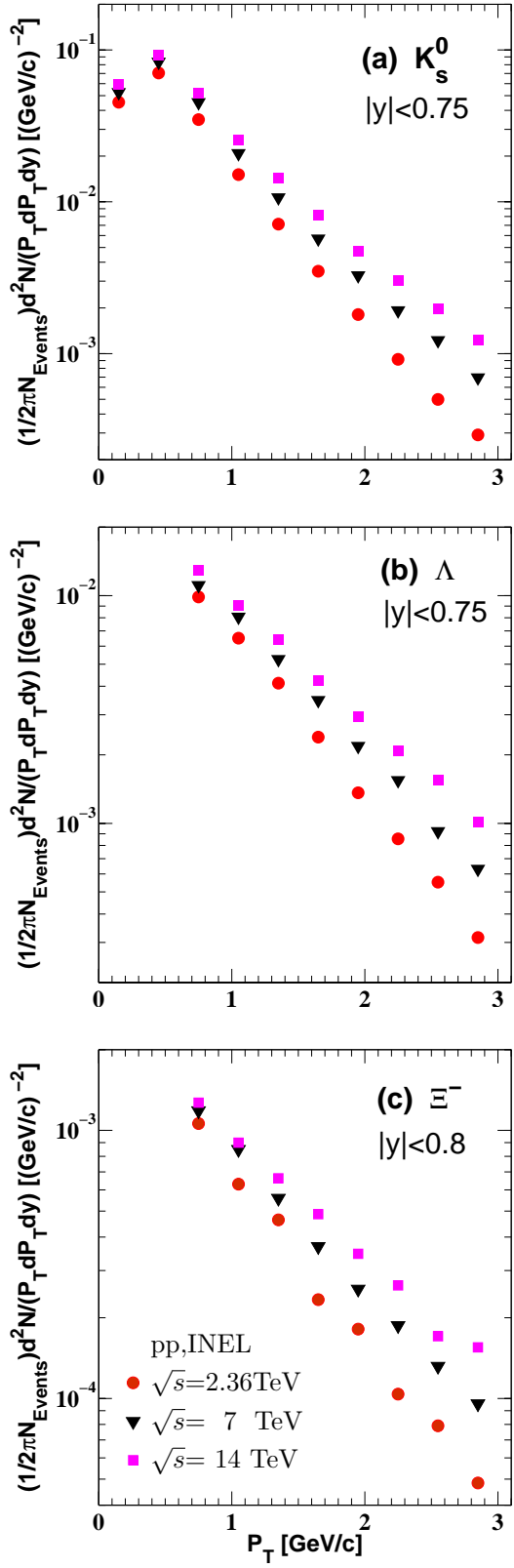}}
\caption{(Color online)~Transverse momentum distributions of
strange particles in INEL pp collisions at $\sqrt s$=2.36, 7, and
14 TeV calculated by PACIAE. Panels (a), (b), and (c) are for
$K_s^0$ ($|y|<$0.75), $\Lambda$ ($|y|<$0.75), and $\Xi^{-}$
($|y|<$0.80), respectively.} \label{predi}
\end{figure}

Similarly, the transverse momentum distribution of $K_s^0$, $\Lambda$, and
$\Xi^-$ in INEL pp collisions at $\sqrt s$=2.36, 7, and 14 TeV were predicted
by the modified PACIAE model and given in Fig.~\ref{predi}. These results are 
prepared to compare with the future experimental measurements at the LHC.

In \cite{alice2,alice3}, the experimental kaon to pion ratios in pp
and/or $p\bar p$ collisions at different reaction energies from
SPS to LHC were plotted in a $K/\pi$ verses $\sqrt s$ figure, although
their experimental details in the measured observable ($K^+/\pi^+$, or 
($K^++K^-$)/($\pi^++\pi ^-$), or $K^0/\pi^0$) and the rapidity acceptance 
were different from each other. These experimental ratios are copied
in Fig.~\ref{kpi}. In this figure, NA49, STAR, ALICE, E735,
and UA5 data were taken from ~\cite{alt}, \cite{star2},
\cite{alice2,alice3}, \cite{alex}, and \cite{alne}, respectively.
In \cite{alice2,alice3}, it was concluded that the ``ratio shows a
slight increase from $\sqrt s$=200 GeV ($K/\pi=0.103\pm0.008$) to
$\sqrt s$=900 GeV ($K/\pi=0.123\pm0.004\pm0.010$), yet consistent
within the error bars. The results at 7 TeV will show whether the
$K/\pi$ ratio keeps rising slowly as a function of $\sqrt s$ or saturates." 
To respond to this challenge the ($K^++K^-$)/($\pi^++\pi^-$) ratio 
was extracted from the above PACIAE model simulations for pp collisions at 
$\sqrt s$=0.2, 0.9, 2.36, 7, and 14 TeV and given in Fig.~\ref{kpi} by red 
triangles. In addition, the default PACIAE and PYTHIA models were used 
to calculate the NA49 ($K^++K^-$)/($\pi^++\pi^-$) ratio in pp collisions at 
158 GeV/c beam momentum. The model parameters employed in these calculations
were the same as the ones used in the above simulations, except $K$=1 was 
assumed because of the low energy.

We find in Fig.~\ref{kpi} that the PACIAE results at $\sqrt s$=17.2, 200 and
900 GeV agree with the NA49, STAR, and ALICE data, respectively, within  
error. Together with the PACIAE results at $\sqrt s$=2.36, 7, and 14
TeV, it seems to indicate that the $K/\pi$ ratio as a function of $\sqrt s$
first increases slightly from $\sqrt s$=0.2 to 0.9 TeV and then 
saturates. Of course, this rough trend of the $K^+/\pi^+$ ratio as a
function of $\sqrt s$ in pp reactions has to be further studied in detail 
both experimentally and theoretically.

\section{Conclusions}
In summary, we used the parton and hadron cascade model PACIAE with the
reduction mechanism of the strange quark suppression and default PYTHIA 
models to systematically investigate the production of strange particles in 
NSD and INEL pp collisions at $\sqrt s$=200 GeV and $\sqrt s$=0.9, 2.36, 7, 
and 14 TeV, respectively. The calculated strangeness transverse momentum
distributions in NSD pp collisions at $\sqrt s$=200 GeV compare well
with the STAR data \cite{star1} and PACIAE is globally better than the 
default PYTHIA model. We also compared the calculated strange particle 
rapidity densities at mid-rapidity and transverse momentum distributions in 
INEL pp collisions at $\sqrt s$=0.9 TeV with the corresponding ALICE data 
\cite{alice1}. The agreement between theory and experiment is good and the 
PACIAE is globally better than the default PYTHIA model. This may indicate 
the significant effect of the parton and hadron rescatterings, as well as the 
reduction mechanism of strange quark suppression. These effects are worth 
investigating separately.

\begin{figure}[h!]
\centering \resizebox{0.43\textwidth}{!}{
\includegraphics{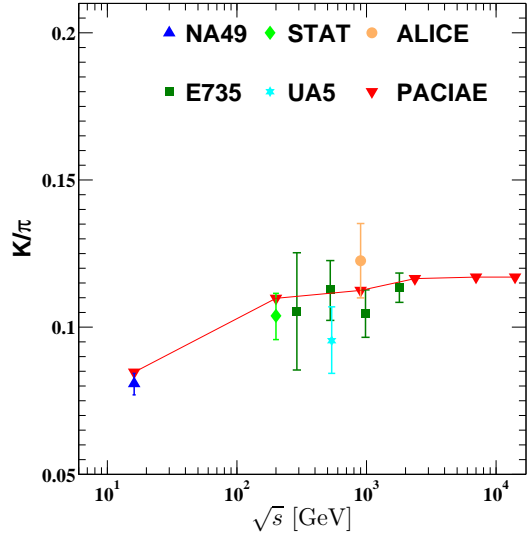}}
\caption{(Color online)~$K/\pi$ ratio as a function of $\sqrt s$
in pp collisions at relativistic energies.} \label{kpi}
\end{figure}

The $K/\pi$ ratios calculated by PACIAE in NSD pp collisions at $\sqrt s$=0.2
TeV and in INEL pp collisions at $\sqrt s$=0.9 TeV agree with the
corresponding STAR \cite{star2} and ALICE data \cite{alice2,alice3}, 
respectively, within error. The PACIAE model result in pp collisions at
158 GeV/c beam momentum is also comparable with NA49 data \cite{alt}.
The rough trend of the $K/\pi$ ratio as function of $\sqrt s$ shown in
Fig.~\ref{kpi} seems to be first increasing moderately, then slightly, and
finally saturating. This trend has to be studied in detail both
experimentally and theoretically.

\begin{figure*}[h!]
\centering \resizebox{0.85\textwidth}{!}{
\includegraphics{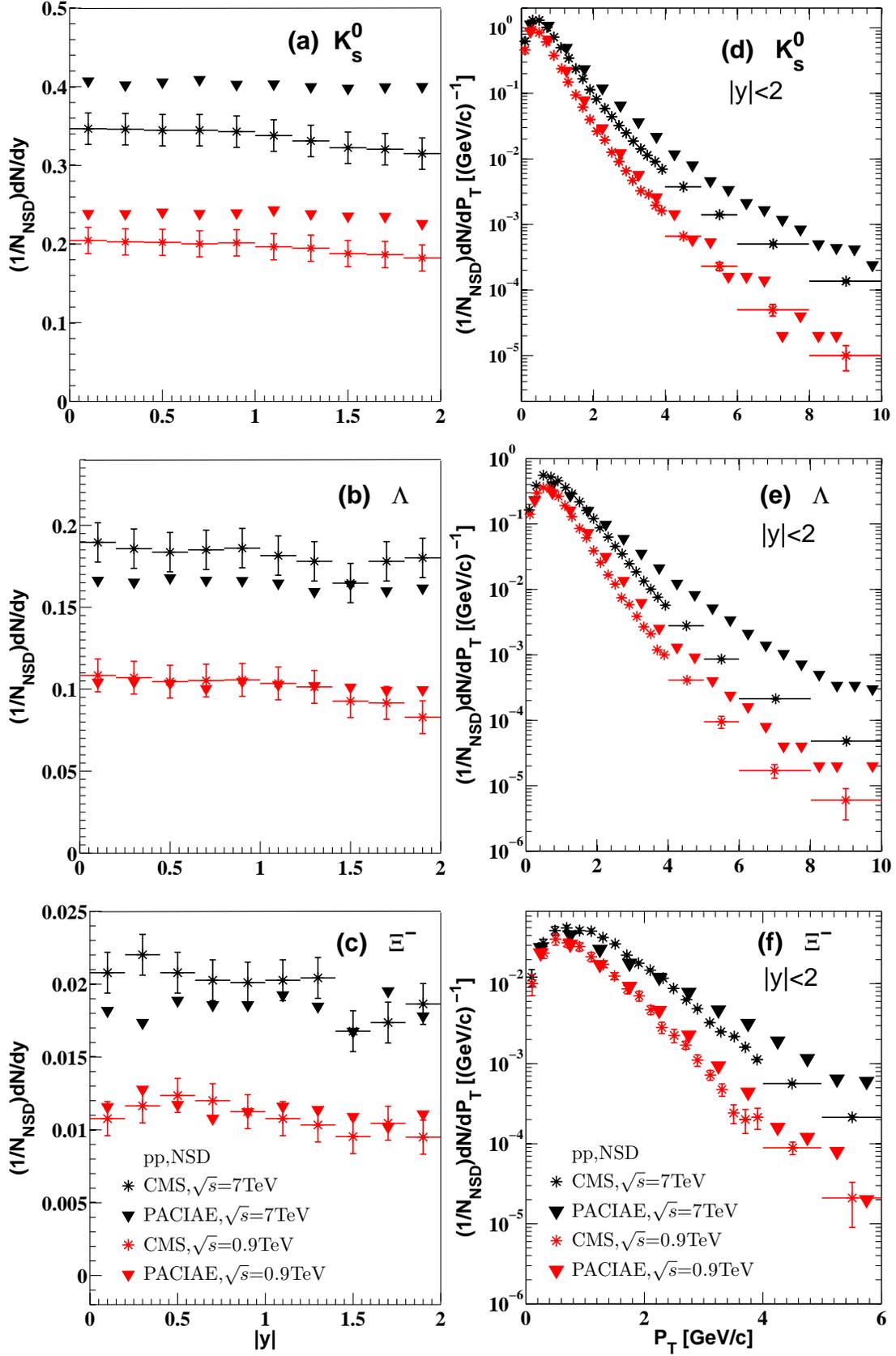}}
\caption{(Color online)~Left panel: the strange particle rapidity
distributions for $K_s^0$ (a), $\Lambda$ (b), and $\Xi^-$ (c) in
NSD pp collisions at $\sqrt s $=0.9 and 7 TeV. Right panel: the
transverse momentum distributions for $K_s^0$ (d), $\Lambda$ (e),
and $\Xi^-$ (f). The CMS data were taken from \cite{cms1}.}
\label{cms}
\end{figure*}

Just before the submission of this paper, we read the paper of \cite{cms1}
and knew that the CMS measured ``$p_T$ distributions are found to differ
substantially from PYTHIA results and the production rates exceed the
predictions by up to a factor of three." Note, here PYTHIA refers to a variety
of PYTHIA tunes quoted in \cite{cms1}. We use the modified PACIAE model to 
calculate the corresponding values. The calculated $dN/dy|_{y\approx 0}$=
0.238, 0.102, and 0.0113 for $K_s^0$, $\Lambda$, and $\Xi^-$ compare well 
with the CMS data of $0.205\pm 0.001\pm 0.015$, $0.108\pm 0.001\pm 0.012$, 
$0.011\pm 0.001\pm 0.001$ \cite{cms1} in NSD pp collisions at $\sqrt s$=0.9 
TeV, respectively. Similarly, the PACIAE results of 0.403, 0.166, and 0.0183 
compare well with the data of $0.346\pm 0.001\pm 0.025$, $0.189\pm 0.001\pm 
0.022$, and $0.021\pm 0.001\pm 0.003$, respectively, in NSD pp collisions at 
$\sqrt s$=7 TeV. The calculated rapidity and transverse momentum 
distributions are compared with the CMS data \cite{cms1} in Fig.~\ref{cms}. 
The agreement is better than the PYTHIA tunes quoted in \cite{cms1}.

\acknowledgments
Financial support from the National Science Foundation of China (10975091,
11075217, 11047142, and 10975062) is acknowledged. BHS would like to thank 
Dr. James Ketudat Cairns for help improving the English.

{}

\begin{table*}
\caption{Strange particles rapidity densities at mid-rapidity ($|y|<$0.5) in
NSD pp collisions at $\sqrt s$=200 GeV. The STAR data were taken from
\cite{star1}. }
\begin{ruledtabular}
\begin{tabular}{ccccc}
 & STAR & PACIAE & PYTHIA\\
\hline
       $K_s^0$& $0.134\pm0.011$ & 0.127 & 0.107\\
       $K^+$  & $0.140\pm0.01$  & 0.135 & 0.112\\
       $K^-$  & $0.137\pm0.009$ & 0.123 & 0.104\\
       $\Lambda$& $0.0385\pm0.0036$ & 0.0360 & 0.0163\\
       $\overline \Lambda$& $0.0351\pm0.0033$ & 0.0350 & 0.0163\\
       $\Xi^-$  & $0.0026\pm0.00092$  & 0.00373 & 0.00106\\
       $\overline\Xi^+$  & $0.0029\pm0.001$ & 0.00364 & 0.00099\\
       $\Omega^-+\overline\Omega^+$  & $0.00034\pm0.00019$ & 0.00026 & 0.00005\\
\end{tabular}
\end{ruledtabular} \label{200}
\end{table*}

\begin{table*}
\caption{Strange particles rapidity densities at mid-rapidity in INEL pp
collisions at $\sqrt s$=900 GeV. The ALICE data were taken from
\cite{alice1}.}
\begin{ruledtabular}
\begin{tabular}{ccccccc}
 &$|y|$&$p_T(GeV/c)$& ALICE & PACIAE & PYTHIA\\
\hline
$K_s^0$& $<0.75$&[0.2-3.0]&$0.184\pm0.0063$ & 0.15 & 0.172\\
$\Phi$& $<0.60$&[0.7-3.0]&$0.021\pm0.005$ & 0.019 & 0.0134\\
$\Lambda$&$<0.75$&[0.6-3.5]& $0.048\pm0.00412$ & 0.043& 0.0227\\
$\overline \Lambda$&$<0.75$&[0.6-3.5]& $0.047\pm0.00539$ & 0.043 & 0.0230\\
$\Xi^-$+$\overline \Xi^+$  & $<0.80$&[0.6-3.0]&$0.0101\pm0.00219$  & 0.0086 & 0.00354\\
\end{tabular}
\label{900}
\end{ruledtabular}
\end{table*}

\end{document}